# Improved formulation for long-duration storage in capacity expansion models using representative periods

*Federico Parolin* [a,b], *Paolo Colbertaldo* [a], *Ruaridh Macdonald* [b]

a. Department of Energy, Politecnico di Milano, Milan, Italy
b. MIT Energy Initiative, Massachusetts Institute of Technology, Cambridge, MA, USA

## Abstract

With the increasing complexity and size of capacity expansion models, temporal aggregation has emerged as a common method to improve computational tractability. However, this approach inherently complicates the inclusion of long-duration storage (LDS) systems, whose operation involves the entire time horizon connecting all time steps. This work presents a detailed investigation of LDS modelling with temporal aggregation. A novel compact formulation is proposed to reduce the number of constraints while effectively tracking the storage content and enforcing limits on the state of charge throughout the entire time horizon. The developed method is compared with two leading state-of-the-art formulations. All three methods are implemented in the Dolphyn capacity expansion model and tested on a case study for the continental United States, considering different configurations in terms of spatial resolutions and representative periods. The performance is assessed with both the commercial solver Gurobi and the open-source solver HiGHS. Results show that the developed compact formulation consistently outperforms the other methods in terms of both runtime (30%-70% faster than other methods) and memory usage (1%-9% lower than other methods).

## Highlights

- A new formulation is developed to model long-duration storage with temporal aggregation
- The proposed method is compared with two state-of-the-art formulations
- All three methods are implemented and tested with the Dolphyn model
- The developed formulation can achieve 30%-70% faster runtimes than other methods
- Memory usage is 1%-9% lower compared to existing formulations

## Nomenclature

| | | |
|---|---|---|
| *Acronyms and abbreviations* | | |
| | CEM | Capacity expansion model |
| | LDS | Long-duration storage |
| | PHS | Pumped-hydro storage |
| | SOC | State of charge |
| *Sets* | | |
| | $h \in H$ | Time steps within entire time horizon |
| | $n \in N$ | Input periods |
| | $t \in T$ | Time steps within period |
| | $w \in W$ | Representative periods |
| *Subscripts and superscripts* | | |
| | inter | Value within representative period |
| | intra | Value between two periods |
| *Decision variables* | | |
| | $C$ | Storage capacity [kWh] |
| | $\dot{E}_{cha}^{w,t}$ | Charge flow within representative period $w$ at times step $t$ [kW] |
| | $\dot{E}_{dis}^{w,t}$ | Discharge flow within representative period $w$ at times step $t$ [kW] |
| | $SOC^h$ | State of charge at time step $h$ [kWh] |
| | $SOC_{inter}^n$ | State of charge at the beginning of period $n$ [kWh] |
| | $SOC_{intra}^{w,t}$ | State of charge within representative period $w$ at times step $t$ [kWh] |
| | $\Delta SOC^w$ | Net state of charge variation across representative period $w$ [kWh] |
| | $\Delta SOC_{max,pos}^w$ | Maximum positive state of charge variation within representative period $w$ [kWh] |
| | $\Delta SOC_{max,neg}^w$ | Maximum negative state of charge variation within representative period $w$ [kWh] |
| *Parameters* | | |
| | $\Delta t$ | Time step duration [h] |
| | $\eta_{cha}$ | Charging efficiency [-] |
| | $\eta_{dis}$ | Discharging efficiency [-] |
| | $\eta_{sdc}$ | Self-discharge efficiency [-] |

## 1. Introduction

Energy storage is essential to enable the penetration of intermittent renewable energy sources and achieve net-zero $CO_2$ emissions [1]. In renewables-dominated systems, multiple storage solutions must operate on different time scales to optimally integrate renewable power generation. While existing battery storage systems are a viable option to balance intra-day generation-demand mismatches, long-duration storage (LDS) systems, such as metal-air batteries, hydrogen storage and pumped-hydro storage (PHS), are needed to compensate for seasonal differences in the availability of renewables [2]. Numerous works have investigated the role of LDS in decarbonisation using linear programming (LP)-based capacity expansion models (CEMs) [2–4]. The validity of these results depends on LDS being correctly modelled in the respective CEMs, particularly when optimising over representative periods rather than the entire timeseries.

Researchers have developed more complex CEMs in order to answer contemporary questions about new energy technologies, policies, and decarbonization strategies. These developments have included introducing additional energy sectors, increasing the geographical resolution, and extending the time horizon to several years or even decades. To allow these models to be solved in a reasonable time, a commonly adopted simplification is to reduce the number of modelled time steps through the aggregation of input time series into representative periods[1] [5]. Representative periods are usually generated by selecting sections of the original full timeseries that have similar properties to several other sections of the timeseries, rather than combining or averaging several sections. The sections of the timeseries that are not selected (i.e., the non-representative periods) are not modelled and are assumed to have identical demand and operations as their representative periods.

Temporal aggregation is a pivotal feature of modern CEMs, and many relevant studies and popular frameworks make extensive use of representative periods [6]. One example is the Net-Zero America study, where Larson et al. used the RIO CEM to analyse the evolution of the US energy system over a time horizon of 30 years. The 30 years were modelled using a sequence of seven optimisations with a five-year time step. Each five-year optimisation is modelled with 41 representative days and including both short- and long-duration storage [7]. The widely employed TIMES framework is also based on representative periods. Simoes et al. used the JRC-EU-TIMES model to investigate the European transition to carbon neutrality, analysing the time frame 2005-2075 with a 5-year time step with 12 representative periods for each time step [8]. Different LDS technologies are included in the analysis, such as hydrogen storage, PHS, and underground thermal storage. Kountouris et al. further investigated the European case study using the Balmorel model [9]. The time horizon 2020-2050 is modelled with a 5-year time step considering 16 representative periods for each modelled year, and including hydrogen storage as LDS option. At the global scale, Löffler et al. developed the OSeMOSYS-based model GENeSYS-MOD to investigate the transition to 2050, using 6 representative periods for each modelled year and including various LDS systems (PHS, hydrogen, and centralised heat) [10].

Including LDS technologies in models with time aggregation is inherently challenging. Representative periods are defined as independent time frames, which share no operating or state variables. They are only linked by capacity variables, emission limits, and other features, which are not time dependent. Each representative period is assumed to have cyclic operation. For storage systems, this implies that the stored energy at the end of each representative period must be equal to that at the beginning of the representative period, and the state of charge (SOC) is not carried forward to the next representative period, whereas connection with other periods is absent. While this assumption proves accurate for short-duration storage systems (e.g., batteries), it fails to accurately represent LDS operation, whose value inherently comes from storing energy over long periods. For LDS to be properly considered, its stored energy must be tracked over the entire time horizon of the model.

Different approaches have been proposed to address this challenge. Gabrielli et al. developed a formulation where the constraints to track the energy stored in a LDS systems are explicitly defined for the entire time horizon, including the non-representative periods. For each time step of the original timeseries, the stored energy is calculated based on the charge/discharge flow of the corresponding time step of the corresponding representative period [11]. The same approach is implemented in the EnergyScope TD CEM [12,13].

While Gabrielli et al.'s method enables explicit tracking of the storage inventory evolution throughout the entire time horizon, it is also possible to use an implicit approach which requires fewer variables and constraints, reducing the complexity of the model. The Dolphyn [14] and GenX [15] CEMs implement implicit tracking, only computing the storage content at the time steps within representative periods and at the beginning of each representative and non-representative period. However, the formulation used does not guarantee that the storage inventory does not exceed the maximum and minimum state of charge within the non-representative periods. Kotzur et al. developed a correct implicit LDS formulation, modelling LDS by considering the superposition of two time layers, one that models the operation within the representative periods, and one that model changes between periods [16]. Dedicated constraints are introduced across the full time horizon to prevent violations of the state of charge limits.

---
[1] In the literature, representative periods are also referred to as typical periods or time slices.

Within this context, this work conducts a detailed investigation of long-duration storage modelling in CEMs. A novel formulation, named Implicit - Min-Max is presented that enables a reduction in the number of constraints while ensuring that no violations of the state of charge limits occur within non-representative periods. The computational performance, in terms of both solution time and memory usage, is evaluated in comparison with the two principal formulations that have been proposed in the literature (Gabrielli et al. [11] and Kotzur et al. [16]). All the three formulations are implemented in the open-source CEM Dolphyn [14], and the computational performance is evaluated using both a commercial (Gurobi [17]) and an open-source (HiGHS [18]) solver. The proposed method is also compared to the original formulation implemented in the Dolphyn and GenX models, evaluating the occurrence of violations of state of charge limits in non-representative periods.

This paper is structure as follows. Section 2 presents the two state-of-the art LDS modelling formulations and introduces the novel compact formulation. In Section 3, the three formulations are tested and their performance is evaluated considering different configurations in terms of spatial resolution and representative periods. Finally, Section 4 summarises the main conclusions and the most relevant insights.

## 2. Long-duration storage formulations

This section presents the two state-of-the-art formulations for modelling LDS with temporal aggregation and introduces a novel approach to reduce the number of constraints. The methods proposed by Gabrielli et al. [11] (named Explicit - Hourly, discussed in Section 2.2) and Kotzur et al. [16] (named Implicit - Hourly, discussed in Section 2.3) are considered as benchmarks for the novel formulation (named Implicit - Min-Max, discussed in Section 2.4). The list of the adopted symbols is available in the Nomenclature section.

The three methods, together with the original Dolphyn/GenX formulation, are conceptually schematised in Figure 1. The figure shows an example of storage content evolution as represented by the different formulations. Continuous lines indicate explicitly tracked SOC variables, while dashed lines represent implicitly tracked SOC. Crosses denote SOC variables defined over individual time steps (i.e., hours), while time steps where a storage balance constraint is defined are highlighted in light-green. The figure considers three periods, 1,2, and 3. Period 1 corresponds to representative period I, which is also used to represent period 2. Period 3 is instead represented by representative period III. The mapping of periods is also represented in the table in the lower-right part of the figure. The different characteristics of the analysed formulations are discussed in the following sections.

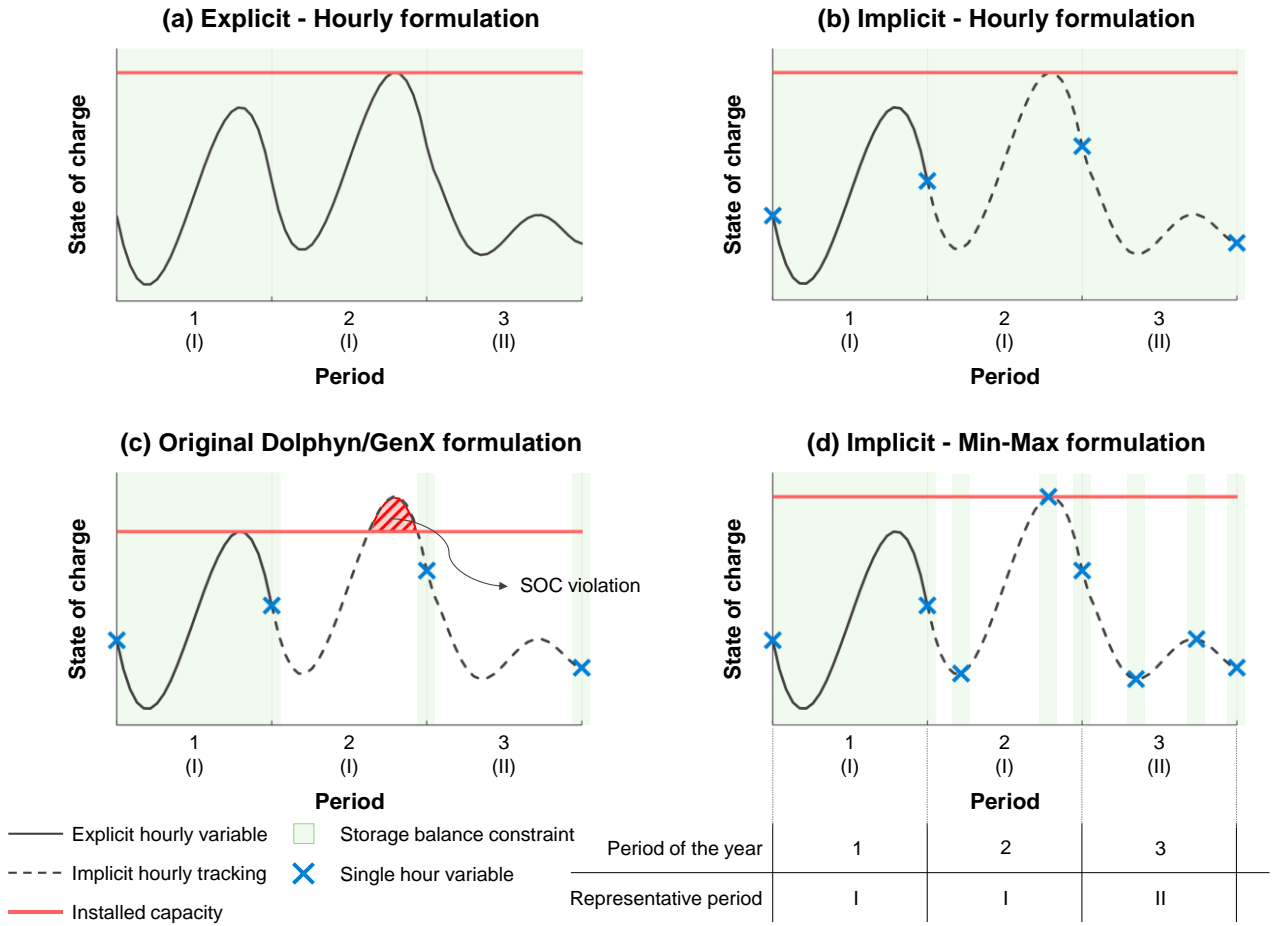

Figure 1. Schematic representation of the analysed long-duration storage (LDS) formulations. Continuous lines indicate explicitly tracked SOC variables, while dashed lines represent implicitly tracked SOC variables. Crosses denote SOC variables defined over individual time steps (hours) Time steps where a storage balance constraint is defined are highlighted in light-green. The x-axis displays periods of the year, with corresponding representative periods in brackets.

## 2.1. Time horizon modelling with temporal aggregation

All three formulations consider a time horizon of $h \in H$ time steps (for a year-long analysis with hourly resolution, $|H| = 8760$ and $h = 1, 2, \ldots, 8760$) organised in $n \in N$ contiguous periods made of $t \in T$ time steps, such that $|H| = |N| \cdot |T|$. As a result of the temporal aggregation, every period of the time horizon $n \in N$ is modelled with a representative period $w \in W$, indicated as $w(n)$, consisting of $t \in T$ time steps (e.g., if $|T| = 24$, representative periods correspond to typical days). Representative periods are properly repeated, such that every original time step $h \in H$ is represented by a time step $t \in T$ within a representative period $w \in W$. To differentiate from the representative periods, the periods of the time horizon $n \in N$ are also referred to as input periods.

## 2.2. Explicit - Hourly bounds formulation

The formulation developed by Gabrielli et al. [11], here named Explicit - Hourly, explicitly tracks the state of charge throughout the time horizon, including non-representative periods, as represented in Figure 1.a. At each time step $h \in H$, the state of charge $SOC^h$ of a generic long-duration storage system is computed taking into account the state of charge at the previous time step and the charge and discharge flows of the corresponding time step $t \in T$ in the corresponding representative period $w(n)$:

$$n \in N, t \in T \qquad SOC^{h=|T| \cdot (n-1)+t+1} = SOC^{h=|T| \cdot (n-1)+t} \cdot (1-\eta_{sdc}) + \dot{E}_{cha}^{w(n),t} \cdot \eta_{cha} \cdot \Delta t - \frac{\dot{E}_{dis}^{w(n),t}}{\eta_{dis}} \cdot \Delta t \qquad (1)$$

where $\eta_{sdc}$ is the self-discharge efficiency, $\eta_{cha}$ is the charge efficiency, $\eta_{dis}$ is the discharge efficiency, and $\Delta t$ is the duration of a time step.

A cyclic condition is introduce to ensure the repeatability of the time horizon (Eq. (2)) and the state of charge is imposed to be greater than or equal to zero and lower than or equal to the installed energy capacity $C$ (Eq. (3)).

$$SOC^{h=1} = SOC^{h=|T| \cdot |N|+1} \qquad (2)$$

$$h \in H \qquad 0 \leq SOC^h \leq C \qquad (3)$$

### 2.3. Implicit - Hourly bounds formulation

The formulation proposed by Kotzur et al. [16], here named Implicit – Hourly (Figure 1.b), tracks the state of charge throughout the time horizon by introducing two time layers. The first layer models the system operation within a representative period and is therefore indicated as an intra-period layer. The second layer models the storage inventory variations between periods and is indicated as inter-period time layer. In this way, the state of charge at a generic time step $h \in H$ (i.e., $SOC^h$) is computed as the superposition of the intra-period and inter-period states of charge, as:

$$n \in N, t \in T \qquad SOC^{h=|T| \cdot (n-1)+t} = SOC_{inter}^n \cdot (1-\eta_{sdc}) + SOC_{intra}^{w(n),t} \qquad (4)$$

where $SOC_{inter}^n$ is the inter-period state of charge and represents the storage inventory at the beginning of the input period $n \in N$, while $SOC_{intra}^{w(n),t}$ is the intra-period state of charge at the time step $t \in T$ within the corresponding representative period $w(n)$.

In this formulation, the intra-period state of charge represents the variation of the storage content with respect to the beginning of the period (i.e., with respect to $SOC_{inter}^n$). Accordingly, the intra-period state of charge at the first time step of each representative period must be equal to zero:

$$w \in W \qquad SOC_{intra}^{w,1} = 0 \qquad (5)$$

In the subsequent time steps, the state of charge is computed considering the charge and discharge flows of the modelled time step in the representative period:

$$w \in W, t \in T\backslash\{1\} \qquad SOC_{intra}^{w,t} = SOC_{intra}^{w,t-1} \cdot (1-\eta_{sdc}) + \dot{E}_{cha}^{w,t} \cdot \eta_{cha} \cdot \Delta t - \frac{\dot{E}_{dis}^{w,t}}{\eta_{dis}} \cdot \Delta t \qquad (6)$$

The inter-period states of charge are then defined as:

$$w \in W, n \in N\backslash\{|N|\} \qquad SOC_{inter}^{n+1} = SOC_{inter}^n + SOC_{intra}^{w(n),t=|T|} \qquad (7)$$

$$SOC_{inter}^{n=1} = SOC_{inter}^{n=|N|} + SOC_{intra}^{w(n),t=1} \qquad (8)$$

where $SOC_{intra}^{w(n),t=|T|}$ corresponds to the net state of charge variation across the representative period $w(n)$.

The constraints on the state of charge limits are ensured by superposing the inter- and intra- states of charge, as:

$$n \in N, t \in T \qquad SOC_{inter}^{n} \cdot (1-\eta_{sdc})^{t} + SOC_{intra}^{w(n),t} \leq C \qquad (9)$$

$$n \in N, t \in T \qquad SOC_{inter}^{n} \cdot (1-\eta_{sdc})^{t} + SOC_{intra}^{w(n),t} \geq 0 \qquad (10)$$

### 2.4. Implicit - Min-Max bounds formulation

This section presents a novel formulation (named Implicit - Min-Max) to model long-duration storage with a lower number of constraints while ensuring that the limits on the state of charge are respected throughout the entire time horizon (Figure 1.d). This approach assumes that each representative period $w \in W$ corresponds to a specific period of the time horizon, denoted as $n(w)$ (as, for example, representative period I corresponds to period 1 in Figure 1). Within this period, the storage content $SOC^{h}$ is equal to the intra-period state of charge:

$$w \in W, t \in T \qquad SOC^{h=|T|\cdot(n(w)-1)+1} = SOC_{intra}^{w,t} \qquad (11)$$

Within each representative period $w \in W$, the intra-period state charge evolution is computed as:

$$w \in W, t \in T\setminus\{1\} \qquad SOC_{intra}^{w,t} = SOC_{intra}^{w,t-1} \cdot (1-\eta_{sdc}) + \dot{E}_{cha}^{w,t} \cdot \eta_{cha} \cdot \Delta t - \frac{\dot{E}_{dis}^{w,t}}{\eta_{dis}} \cdot \Delta t \qquad (12)$$

Since the intra-period state of charge represents a real storage content, it must always be positive and not exceed the available capacity:

$$w \in W, t \in T \qquad 0 \leq SOC_{intra}^{w,t} \leq C \qquad (13)$$

The inter-period state of charge $SOC_{inter}^{n}$ is used to track the storage inventory at the beginning of input periods $n \in N$. In addition, the variable $\Delta SOC^{w}$ is introduced to compute the net variation in the storage inventory across a representative period. The inter-period storage balance is then defined as:

$$w \in W, n \in N\setminus\{|N|\} \qquad SOC_{inter}^{n+1} = SOC_{inter}^{n} + \Delta SOC^{w(n)} \qquad (14)$$

$$SOC_{inter}^{n=1} = SOC_{inter}^{n=|N|} + \Delta SOC^{w(n=|N|)} \qquad (15)$$

where, similar to the other methods, a cyclic condition on the time horizon is imposed.
In each input period $n \in N$, the inter-period state of charge and the intra-period state of charge at the last time step of the corresponding representative period are connected as:

$$w \in W, n \in N \qquad SOC_{inter}^{n} = SOC_{intra}^{w(n),t=|T|} - \Delta SOC^{w(n)} \qquad (16)$$

At the beginning of each representative period, the intra-period state of charge is computed from the inter-period state of charge of the corresponding input period $n(w)$ as:

$$w \in W \qquad SOC_{intra}^{w,t=1} = SOC_{inter}^{n(w)} \cdot (1-\eta_{sdc}) + \dot{E}_{cha}^{w,t=1} \cdot \eta_{cha} \cdot \Delta t - \frac{\dot{E}_{dis}^{w,t=1}}{\eta_{dis}} \cdot \Delta t \qquad (17)$$

The full state of charge evolution is thus built as:

$$n \in N \qquad SOC^{h=|T|\cdot(n-1)+1} = SOC_{inter}^{n} \cdot (1-\eta_{sdc}) + \dot{E}_{cha}^{w(n),t=1} \cdot \eta_{cha} \cdot \Delta t - \frac{\dot{E}_{dis}^{w(n),t=1}}{\eta_{dis}} \cdot \Delta t \qquad (18)$$

$$n \in N, t \in T\setminus\{1\} \qquad SOC^{h=|T|\cdot(n-1)+t} = SOC^{h=|T|\cdot(n-1)+t-1} \cdot (1-\eta_{sdc}) + \dot{E}_{cha}^{w(n),t} \cdot \eta_{cha} - \frac{\dot{E}_{dis}^{w(n),t}}{\eta_{dis}} \qquad (19)$$

In the GenX [15] and Dolphyn [14] models, the set of equations (11)-(16) is completed by bounding the inter-period states of charge to be positive and less than the installed energy capacity. However, such an approach does not guarantee that the state of charge limits are not exceeded within non-representative periods (Figure 1.c). To address this issue, an additional set of constraints is here introduced. Two auxiliary variables $\Delta SOC_{max,pos}^{w}$ and $\Delta SOC_{max,neg}^{w}$ are defined to represent the maximum positive and negative storage content variations within a representative period. These are extracted as:

$$w \in W, t \in T\setminus\{1\} \qquad \Delta SOC_{max,pos}^{w} \geq SOC_{intra}^{w,t} - SOC_{intra}^{w,t=1} \qquad (20)$$

$$w \in W, t \in T\setminus\{1\} \qquad \Delta SOC_{max,neg}^{w} \leq SOC_{intra}^{w,t} - SOC_{intra}^{w,t=1} \qquad (21)$$

For every input period $n \in N$, the maximum state of charge is computed and constrained to be less than or equal to the installed energy capacity as:

$$n \in N \qquad SOC_{inter}^{n} \cdot (1-\eta_{sdc}) + \dot{E}_{cha}^{w(n),t=1} \cdot \eta_{cha} \cdot \Delta t - \frac{\dot{E}_{dis}^{w(n),t=1}}{\eta_{dis}} \cdot \Delta t + \Delta SOC_{max,pos}^{w} \leq C \qquad (22)$$

Similarly, the minimum state of charge is imposed to be positive in every period of the time horizon:

$$n \in N \qquad SOC_{inter}^{n} \cdot (1-\eta_{sdc}) + \dot{E}_{cha}^{w(n),t=1} \cdot \eta_{cha} \cdot \Delta t - \frac{\dot{E}_{dis}^{w(n),t=1}}{\eta_{dis}} \cdot \Delta t + \Delta SOC_{max,neg}^{w} \geq 0 \qquad (23)$$

Compared to the Implicit - Hourly method, the proposed compact formulation enables ensuring that the state of charge limits are not violated throughout the full time horizon by defining two constraints instead of |T| for every input period $n \in N$. Such a difference can be graphically observed in Figure 1, looking at the time steps highlighted in light-green for the different approaches.

This method has been independently developed, building upon the formulation implemented in the GenX and Dolphyn models. A similar approach that extracts the maximum variations of the intra-period state of charge was also presented in Appendix B of Ref. [16]. However, the methodology proposed here is based on a different representation and on distinct assumptions. In addition, this work advances the understanding of the strengths and weaknesses of the various formulation by providing a comprehensive assessment of each method in the following section. Finally, as the Implicit - Min-Max method resolves the state of charge violations present in the original (v0.2.3) Dolphyn formulation (see Section 3.2), it has been integrated into the update Dolphyn version (v0.3.0).

## 3. Results and discussion

The three formulations presented in Section 2 are implemented in the Dolphyn model to evaluate their performance. Section 3.1 outlines the case study used for the evaluations. Section 3.2 compares the Implicit - Min-Max method with the original Dolphyn and GenX implementation to show that violations of the state of charge limits occur if dedicated constraints are not introduced. Finally, Section 0 provides a performance assessment of the three methods.

### 3.1. Numerical experiments and computational setup

The alternative LDS formulations are tested on models of the continental United States made up of 4, 16, and 64 zones, based on the spatial resolution used in the Integrated Planning Model (IPM) of the Environmental Protection Agency (EPA) [19], as schematised in Figure 2. The selection of zones follows the number order of the IPM zones, starting from the Electric Reliability Council of Texas (ERCOT) and Midcontinent Independent System Operator (MISO) regions. Temporal aggregation is performed using k-means clustering with 26, 52, and 104 representative days on a year-long time horizon, leading to a total of 9 configurations. The analysis considers the electricity and hydrogen sectors, and data are taken from Ref. [20]. The input time series are renewables availability (solar photovoltaic, onshore wind, offshore wind, Run-of-River hydro), fuel prices (natural gas, uranium), electricity demand, and hydrogen demand. Hydrogen storage is modelled as LDS technology, considering one system for each zone.

For each configuration, the optimisation problem is implemented in Dolphyn, using Julia 1.10.1 [21] and JuMP 1.20.0 [22] and solved with either Gurobi 11.02 [17] or HiGHS 1.7 [18]. All numerical experiments are run in the MIT SuperCloud high-performance computing environment [23] with Intel Xeon Platinum 8260 2.40 GHz processors and 192 GB RAM per node. All data and code used in the numerical experiments are available at https://zenodo.org/doi/10.5281/zenodo.13843899 [24].

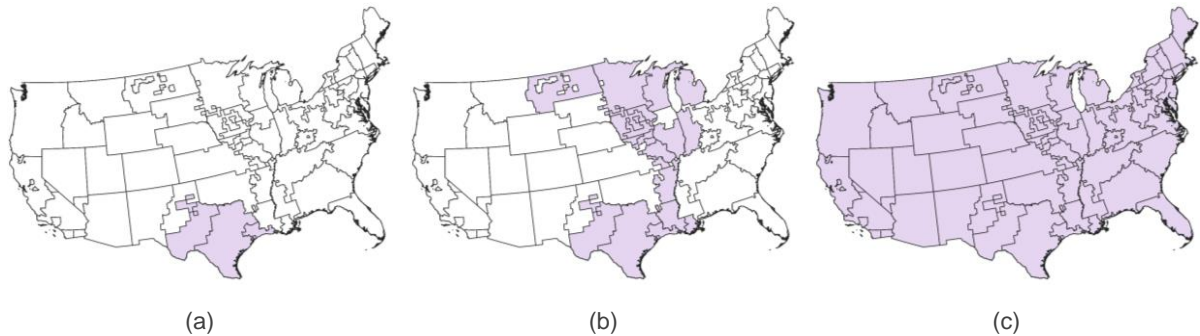

Figure 2. Spatial configurations considered in the numerical experiments: 4 zones (a), 16 zones (b), and 64 zones (c).

### 3.2. Analysis of capacity violations without capacity constraints in non-representative periods

The proposed Implicit - Min-Max method is tested against the original LDS formulation implemented in the Dolphyn and GenX models, which lacks dedicated constraints to limit storage content during non-representative periods. The two methods are compared in terms of runtime, difference in installed LDS energy capacity, and number of violations of the state of charge limits. A violation occurs when the state of charge (computed as in Eqs. (18)-(19)) is either negative or exceeds the installed energy capacity of the storage. The tests are performed with Gurobi (crossover deactivated) for the 16-zone configuration, using 26, 52, and 104 representative days. Since violations are likely to occur less frequently over longer representative periods, additional tests are conducted considering 4, 8, and 15 representative weeks.

Results are summarised in Table 1. The original implementation shows a significant number of violations, ranging between 991-2484 over a total of 8760 time steps, even when representative weeks are used instead of days. This corresponds to 12-34% of the occurrences among time steps of non-representative periods. Figure 2 provides an example of this behaviour, showing the storage content evolution of hydrogen LDS in the West ERCOT zone over 9 periods of the year from the 15-week case for both the original (Figure 2.a) and the Implicit - Min-Max (Figure 2.b) formulations. This figure includes representative periods (weeks 7, 11, and 15, highlighted in light blue) and non-representative periods. The original formulation ensures that the storage content stays within limits during representative periods and at the start of each period. However, in periods 8 to 10, where the initial storage inventory is near the maximum, the state of charge exceeds the installed capacity, and in period 12, it becomes negative. This issue is not observed with the Implicit - Min-Max formulation (Figure 2.b), which ensures that the state of charge remains within the limits.

The absence of dedicated bounding constraints in the original formulation allows LDS systems to be operated more flexibly. As a result, the system tends to make greater use of this option, and the original formulation generally leads to higher installed energy capacity. This effect becomes less pronounced when weeks are used as representative periods instead of days, and is even reversed in the 15-week configuration. Overall, the two methods are comparable performance-wise, as runtime differences are limited and decrease when increasing the number of representative periods.

Table 1. Result comparison between the Implicit - Min-Max and the original Dolphyn and GenX formulations.

| Number of representative periods | Implicit - Min-Max formulation | Original formulation | | |
|---|---|---|---|---|
| | Runtime [s] | Runtime [s] | Average number of violations per LDS system | Installed energy capacity difference |
| 26 days | 39.96 | 25.94 | 991 | +287% |
| 52 days | 101.11 | 82.36 | 1204 | +136% |
| 104 days | 251.73 | 240.78 | 1058 | +156% |
| 4 weeks | 25.46 | 21.27 | 0* | - |
| 8 weeks | 66.29 | 49.59 | 2484 | +59% |
| 15 weeks | 236.44 | 175.50 | 1210 | -68% |

*Optimisation results do not feature LDS storage capacity in this configuration

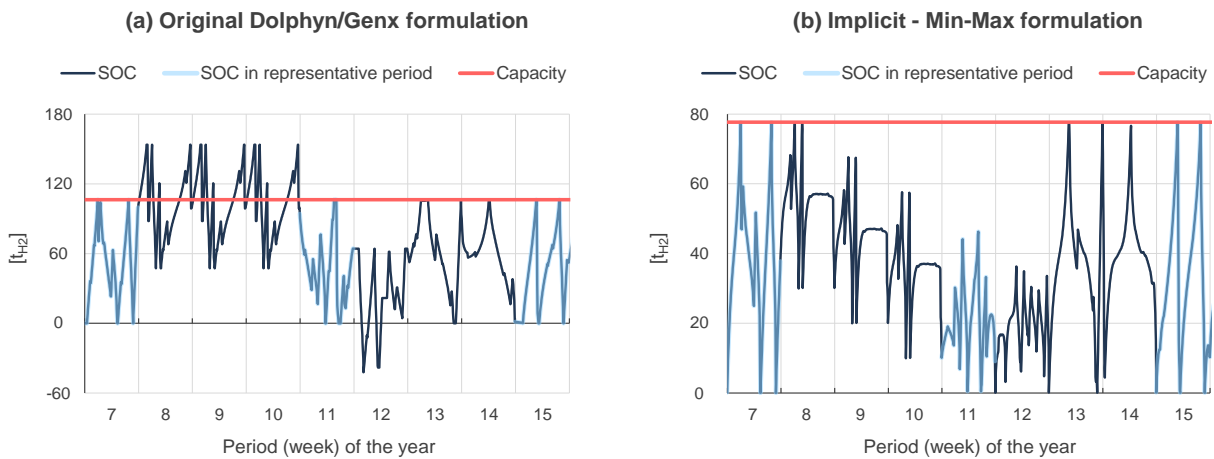

Figure 3. Storage content evolution for hydrogen LDS in the West ERCOT region in the 15-week case as obtained with the original Dolphyn/GenX implementation (a) and with the Implicit - Min-Max formulation (b).

### 3.3. Long-duration storage formulations comparison

This section provides a performance assessment of the three LDS formulations. The methods are tested on the 9 configurations introduced in Section 3.1, first using the commercial solver Gurobi (Section 3.3.1), and then the open-source solver HiGHS (Section 3.3.2). For both solvers, optimisations are performed with and without crossover, and imposing a runtime limit of 10800 seconds. The assessment is concluded by analysing memory usage for the different formulations (Section 3.3.3).

#### 3.3.1 Performance assessment with Gurobi

Table 2 shows the computational performance comparison using Gurobi with crossover disabled, which corresponds to the most commonly used solver-setting combination in energy system modelling. Benefitting from the lowest number of constraints, the Implicit - Min-Max formulation provides the best performance across all configurations. However, the runtime trend does not always correlate with the number of constraints, as the Explicit - Hourly method occasionally outperforms the Implicit - Hourly formulation. This can be attributed to Gurobi's presolve, which, in some cases, is able to generate a tighter model with the Explicit - Hourly method. As an example, Table 3 shows the number of constraints before and after presolve for the 64-zone, 104-day configuration, for which the Explicit - Hourly formulation results having a lower number of constraints than the Implicit - Hourly.

In terms of relative performance gains, the Implicit - Min-Max method is on average 44% faster than the Explicit - Hourly and 34% faster than the Implicit - Hourly. The greatest reductions in runtime achieved are in the order of 70%. These are observed in the 16-zone, 104-day configuration in comparison with the Implicit - Hourly method and in the 64-zone, 52-day configuration in comparison with the Explicit - Hourly method.

Table 2. Runtime (in seconds) for the three LDS formulations using Gurobi with crossover disabled. The best-performing formulation for each configuration is highlighted in bold.

| Formulation | Zones | Representative days | | |
|---|---|---|---|---|
| | | 26 | 52 | 104 |
| Explicit - Hourly | 4 | 10.67 | 17.18 | 50.00 |
| | 16 | 81.26 | 136.79 | 511.21 |
| | 64 | 1144.36 | 2923.09 | 2329.92 |
| Implicit - Hourly | 4 | 7.39 | 12.59 | 28.50 |
| | 16 | 113.68 | 185.17 | 814.79 |
| | 64 | 775.48 | 920.02 | 3178.17 |
| Implicit - Min-Max | 4 | **5.50** | **9.15** | **21.31** |
| | 16 | **36.96** | **101.11** | **251.73** |
| | 64 | **665.49** | **882.73** | **2226.53** |

Table 3. Role of Gurobi's presolve: number of constraints before and after presolve for the 64-zone 104-day configuration.

| Number of constraints | Explicit - Hourly | Implicit - Hourly | Implicit - Min-Max |
|---|---|---|---|
| Before presolve | 22,074,340 | 21,696,612 | 21,117,988 |
| After presolve | 8,930,777 | 9,006,105 | 8,524,207 |

The runs are repeated with crossover enabled, and results are reported in Table 4. Although the Implicit - Min-Max formulation continues to deliver the best performance in the majority of configurations, the trend is less consistent with this setting, as presolve plays an even more significant role. Runtimes are fairly similar for the 4-zone, 104-day configuration, for which the Explicit - Hourly formulation offers the best performance. On the other hand, the largest performance gains are achieved with the Implicit - Min-Max method in the 26-day cases, for which the Implicit - Hourly is the worst performing formulation.

Table 4. Runtime (in seconds) for the three LDS formulations using Gurobi with crossover enabled. The best-performing formulation for each configuration is highlighted in bold.

| Formulation | Zones | Representative days | | |
|---|---|---|---|---|
| | | 26 | 52 | 104 |
| Explicit - Hourly | 4 | 14.39 | **34.20** | 51.25 |
| | 16 | 120.00 | 2382.79 | 51671.58 |
| | 64 | > 10800 | > 10800 | > 10800 |
| Implicit - Hourly | 4 | 71.70 | 70.43 | 55.50 |
| | 16 | 340.09 | 1740.67 | **3443.97** |
| | 64 | > 10800 | > 10800 | > 10800 |
| Implicit - Min-Max | 4 | **7.65** | 62.77 | 52.01 |
| | 16 | **49.44** | 1607.59 | 5495.51 |
| | 64 | > 10800 | > 10800 | > 10800 |

For completeness, simulations for the 16-zone system are repeated using representative weeks. To maintain a comparable number of time steps to the representative day configurations, 4, 8, and 15 representative weeks are considered. As Table 5 shows, the Implicit - Min-Max method outperforms the other formulations across all configurations. In the 8-week case, the three methods exhibit similar performance, with runtime reductions of 3%-6%. Instead, the Implicit - Min-Max method is 56%-72% and 23%-67% faster than the others in the 4-week and 15-week configurations, respectively. On average, the performance gain is more moderate if compared to the analogous 16-zone configurations using representative days, for which the Implicit - Min-Max formulation is on average 52% faster than the others. Nevertheless, the improvement remains significant, with an average runtime reduction of 28%.

Table 5. Runtime (in seconds) for the 16-zone system with representative weeks (using Gurobi with crossover disabled). The best-performing formulation for each configuration is highlighted in bold.

| Formulation | Representative weeks | | |
|---|---|---|---|
| | 4 | 8 | 15 |
| Explicit - Hourly | 90.50 | 70.15 | 710.05 |
| Implicit - Hourly | 57.46 | 68.36 | 306.86 |
| Implicit - Min-Max | **25.46** | **66.29** | **236.44** |

### 3.3.2 Performance assessment with HiGHS

Results obtained with the open-source solver HiGHS with crossover disabled are presented in Table 6. Due to the less advanced presolve in HiGHS, runtime is proportional to the number of constraints. Accordingly, the Implicit - Min-Max method consistently outperforms the other formulations across all configurations, while the Explicit - Hourly method always provides the worst performance. In contrast to the Gurobi assessment (see Table 2), the relative performance gain tends to decrease as the number of representative days increases. Nevertheless, the Implicit - Min-Max formtulation remains 10% to 30% faster than the Implicit - Hourly method and 17% to 49% faster than the Explicit - Hourly.

The same trend is observed if crossover is enabled, as shown in Table 7. In this case, the Implicit - Min-Max is the only formulation capable of solving a 16-zone configuration within the imposed runtime limit. Overall, computational times are significantly longer with HiGHS compared to Gurobi.

Table 6. Runtime (in seconds) for the three LDS formulations using HiGHS with crossover disabled. The best-performing formulation for each configuration is highlighted in bold.

| Formulation | Zones | Representative days | | |
|---|---|---|---|---|
| | | 26 | 52 | 104 |
| **Explicit - Hourly** | 4 | 556.15 | 1337.26 | 4722.38 |
| | 16 | 8314.49 | > 10800 | > 10800 |
| | 64 | > 10800 | > 10800 | > 10800 |
| **Implicit - Hourly** | 4 | 433.01 | 1170.99 | 4307.98 |
| | 16 | 5993.58 | > 10800 | > 10800 |
| | 64 | > 10800 | > 10800 | > 10800 |
| **Implicit - Min-Max** | 4 | 298.59 | 1043.16 | 3917.67 |
| | 16 | 4200.61 | > 10800 | > 10800 |
| | 64 | > 10800 | > 10800 | > 10800 |

Table 7. Runtime (in seconds) for the three LDS formulations using HiGHS with crossover enabled. The best-performing formulation for each configuration is highlighted in bold.

| Formulation | Zones | Representative days | | |
|---|---|---|---|---|
| | | 26 | 52 | 104 |
| **Explicit - Hourly** | 4 | 831.40 | 2361.58 | 7582.32 |
| | 16 | > 10800 | > 10800 | > 10800 |
| | 64 | > 10800 | > 10800 | > 10800 |
| **Implicit - Hourly** | 4 | 494.60 | 1842.97 | 6330.61 |
| | 16 | > 10800 | > 10800 | > 10800 |
| | 64 | > 10800 | > 10800 | > 10800 |
| **Implicit - Min-Max** | 4 | 385.86 | 1646.85 | 6163.30 |
| | 16 | 4786.52 | > 10800 | > 10800 |
| | 64 | > 10800 | > 10800 | > 10800 |

### 3.3.3 Memory usage

Table 8 shows memory usage for the three LDS formulations across the analysed configurations. As expected, the Implicit - Min-Max method has the lowest memory requirements due to its reduced number of constraints, using 1%-9% less memory than the Explicit - Hourly and 0.7%-6% less than the Implicit - Hourly. While the absolute memory savings are not considerable, with a maximum reduction in the range of 2-3 GB, this difference could become more pronounced in larger-scale systems where memory usage is a major factor in problem tractability.

*Table 8. Memory usage (in GB) for the three LDS formulations. The best-performing formulation for each configuration is highlighted in bold.*

| Formulation | Zones | Representative days | | |
|---|---|---|---|---|
| | | 26 | 52 | 104 |
| **Explicit - Hourly** | 4 | 6.71 | 7.81 | 9.99 |
| | 16 | 11.47 | 16.57 | 26.74 |
| | 64 | 37.81 | 66.03 | 122.95 |
| **Implicit - Hourly** | 4 | 6.68 | 7.80 | 9.98 |
| | 16 | 11.25 | 16.42 | 26.67 |
| | 64 | 36.55 | 65.29 | 122.67 |
| **Implicit - Min-Max** | 4 | **6.56** | **7.67** | **9.90** |
| | 16 | **10.69** | **15.88** | **26.29** |
| | 64 | **34.37** | **63.27** | **121.08** |

## 4. Conclusions

This work performed a detailed investigation of long-duration storage (LDS) modelling with temporal aggregation. A novel compact formulation was introduced and compared with the two main state-of-the art methods developed by Gabrielli et al. and Kotzur et al.. The novel formulation, named Implicit - Min-Max, tracks the storage content over the entire time horizon and enforces the storage content limits in both the representative and non-representative periods, while requiring a lower number of constraints compared to existing methods. All three formulations were implemented in the Dolphyn capacity expansion model and tested on a case study for the continental United States, considering multiple configurations in terms of spatial resolution and number of representative periods.

The Implicit - Min-Max formulation was first tested against the original implementation of the Dolphyn and GenX models to demonstrate the necessity of dedicated constraints for ensuring that storage content limits are respected within non-representative periods. The original formulation exhibited a significant number of violations, with the state of charge either exceeding the installed capacity or becoming negative in 991 to 2484 time steps out of a total of 8760 per LDS system, on average. In addition, the installed capacity is often overestimated, as the systems tends to make greater use of LDS systems, which offer greater advantages due to the additional flexibility that results from the absence of bounding constraints in non-representative periods. The proposed formulation eliminates all violations, with minimal impact on the computational performance.

The performance of the three formulations was evaluated using both the commercial solver Gurobi and the open-source solver HiGHS. Results show that the Implicit - Min-Max formulation introduced in this work consistently achieves the best runtime. When Gurobi is used with crossover disabled, which is common for large energy system models, runtime can be decreased by up to 30%-70% compared to the other methods. In addition, the developed approach demonstrates the lowest memory usage, with savings ranging from 1% to 9%. These improvements make the developed method particularly well-suited for macro-scale capacity expansion models, which heavily rely on temporal aggregation and for which scalability and tractability are key concerns.

High Perform. Extrem. Comput. Conf. (2018) 1–6.

# Acknowledgements

The authors thankfully acknowledge the Progetto Rocca program for supporting this collaboration between MIT and Politecnico di Milano. F. Parolin gratefully acknowledges the MIT Energy Initiative for their hospitality and support.